\magnification=\magstep1


\newbox\SlashedBox
\def\slashed#1{\setbox\SlashedBox=\hbox{#1}
\hbox to 0pt{\hbox to 1\wd\SlashedBox{\hfil/\hfil}\hss}{#1}}
\def\hboxtosizeof#1#2{\setbox\SlashedBox=\hbox{#1}
\hbox to 1\wd\SlashedBox{#2}}

\def\mathslashed#1{\setbox\SlashedBox=\hbox{$#1$}
\hbox to 0pt{\hbox to 1\wd\SlashedBox{\hfil/\hfil}\hss}#1}

\def\ifsmall{\iffalse}  
\def\titlepagefont{}  

\def\DefineTeXgraphics{%
\special{ps::[global] /TeXgraphics { } def}}  

\def\today{\ifcase\month\or January\or February\or March\or April\or May
\or June\or July\or August\or September\or October\or November\or
December\fi\space\number\day, \number\year}
\def\eatPrefix19{}
\def\Year{\expandafter\eatPrefix\the\year}
\newcount\hours \newcount\minutes
\def\monthname{\ifcase\month\or
January\or February\or March\or April\or May\or June\or July\or
August\or September\or October\or November\or December\fi}
\def\shortmonthname{\ifcase\month\or
Jan\or Feb\or Mar\or Apr\or May\or Jun\or Jul\or
Aug\or Sep\or Oct\or Nov\or Dec\fi}

\def\TimeStamp{\hours\the\time\divide\hours by60%
\minutes -\the\time\divide\minutes by60\multiply\minutes by60%
\advance\minutes by\the\time%
${\rm \shortmonthname}\cdot\if\day<10{}0\fi\the\day\cdot\the\year%
\qquad\the\hours:\if\minutes<10{}0\fi\the\minutes$}







\newif\ifdraftmode
\newif\ifleftlabels  

\def\nolabels{\def\wrlabeL##1{}\def\eqlabeL##1{}\def\reflabeL##1{}}
\def\writelabels{\def\wrlabeL##1{\leavevmode\vadjust{\rlap{\smash%
{\line{{\escapechar=` \hfill\rlap{\sevenrm\hskip.03in\string##1}}}}}}}%
\def\eqlabeL##1{{\escapechar-1\rlap{\sevenrm\hskip.05in\string##1}}}%
\def\reflabeL##1{\noexpand\rlap{\noexpand\sevenrm[\string##1]}}}
\def\writeleftlabels{\def\wrlabeL##1{\leavevmode\vadjust{\rlap{\smash%
{\line{{\escapechar=` \hfill\rlap{\sevenrm\hskip.03in\string##1}}}}}}}%
\def\eqlabeL##1{{\escapechar-1%
\rlap{\sixrm\hskip.05in\string##1}%
\llap{\sevenrm\string##1\hskip.03in\hbox to \hsize{}}}}%
\def\reflabeL##1{\noexpand\rlap{\noexpand\sevenrm[\string##1]}}}
\nolabels

\newdimen\fullhsize
\newdimen\hstitle
\hstitle=\hsize 
\newdimen\hsbody
\hsbody=\hsize 
\newdimen\hbodyoffset
\hbodyoffset=\hoffset 
\newbox\leftpage
\def\abstract#1{#1}
\def\rotated{\special{ps: landscape}
\magnification=1200  
\baselineskip=14pt
\global\hstitle=9truein\global\hsbody=4.75truein
\global\vsize=7truein\global\voffset=-.31truein
\global\hoffset=-0.54in\global\hbodyoffset=-.54truein
\global\fullhsize=10truein
\def\DefineTeXgraphics{%
\special{ps::[global]
/TeXgraphics {currentpoint translate 0.7 0.7 scale
              -80 0.72 mul -1000 0.72 mul translate} def}}
\let\lr=L
\def\ifsmall{\iftrue}
\def\titlepagefont{\twelvepoint}
\trueseventeenpoint
\def\almostshipout##1{\if L\lr \count1=1
      \global\setbox\leftpage=##1 \global\let\lr=R
   \else \count1=2
      \shipout\vbox{\hbox to\fullhsize{\box\leftpage\hfil##1}}
      \global\let\lr=L\fi}

\output={\ifnum\count0=1 
 \shipout\vbox{\hbox to \fullhsize{\hfill\pagebody\hfill}}\advancepageno
 \else
 \almostshipout{\leftline{\vbox{\pagebody\makefootline}}}\advancepageno
 \fi}

\def\abstract##1{{\leftskip=1.5in\rightskip=1.5in ##1\par}} }

\def\linemessage#1{\immediate\write16{#1}}

\global\newcount\secno \global\secno=0
\global\newcount\appno \global\appno=0
\global\newcount\meqno \global\meqno=1
\global\newcount\subsecno \global\subsecno=0
\global\newcount\figno \global\figno=0

\newif\ifAnyCounterChanged
\let\terminator=\relax
\def\normalize#1{\ifx#1\terminator\let\next=\relax\else%
\if#1i\aftergroup i\else\if#1v\aftergroup v\else\if#1x\aftergroup x%
\else\if#1l\aftergroup l\else\if#1c\aftergroup c\else%
\if#1m\aftergroup m\else%
\if#1I\aftergroup I\else\if#1V\aftergroup V\else\if#1X\aftergroup X%
\else\if#1L\aftergroup L\else\if#1C\aftergroup C\else%
\if#1M\aftergroup M\else\aftergroup#1\fi\fi\fi\fi\fi\fi\fi\fi\fi\fi\fi\fi%
\let\next=\normalize\fi%
\next}
\def\makeNormal#1#2{\def\doNormalDef{\edef#1}\begingroup%
\aftergroup\doNormalDef\aftergroup{\normalize#2\terminator\aftergroup}%
\endgroup}

\def\warnIfChanged#1#2{%
\ifundef#1
\else\begingroup%
\edef\oldDefinitionOfCounter{#1}\edef\newDefinitionOfCounter{#2}%
\ifx\oldDefinitionOfCounter\newDefinitionOfCounter%
\else%
\linemessage{Warning: definition of \noexpand#1 has changed.}%
\global\AnyCounterChangedtrue\fi\endgroup\fi}

\def\Section#1{\global\advance\secno by1\relax\global\meqno=1%
\global\subsecno=0%
\bigbreak\bigskip
\centerline{\twelvepoint \bf %
\the\secno. #1}%
\par\nobreak\medskip\nobreak}
\def\tagsection#1{%
\warnIfChanged#1{\the\secno}%
\xdef#1{\the\secno}%
\ifWritingAuxFile\immediate\write\auxfile{\noexpand\xdef\noexpand#1{#1}}\fi%
}
\def\section{\Section}
\def\Subsection#1{\global\advance\subsecno by1\relax\medskip %
\leftline{\bf\the\secno.\the\subsecno\ #1}%
\par\nobreak\smallskip\nobreak}
\def\tagsubsection#1{%
\warnIfChanged#1{\the\secno.\the\subsecno}%
\xdef#1{\the\secno.\the\subsecno}%
\ifWritingAuxFile\immediate\write\auxfile{\noexpand\xdef\noexpand#1{#1}}\fi%
}

\def\subsection{\Subsection}

\def\romappno{\uppercase\expandafter{\romannumeral\appno}}
\def\makeNormalizedRomappno{%
\expandafter\makeNormal\expandafter\normalizedromappno%
\expandafter{\romannumeral\appno}%
\edef\normalizedromappno{\uppercase{\normalizedromappno}}}
\def\Appendix#1{\global\advance\appno by1\relax\global\meqno=1\global\secno=0%
\global\subsecno=0%
\bigbreak\bigskip
\centerline{\twelvepoint \bf Appendix %
\romappno. #1}%
\par\nobreak\medskip\nobreak}
\def\tagappendix#1{\makeNormalizedRomappno%
\warnIfChanged#1{\normalizedromappno}%
\xdef#1{\normalizedromappno}%
\ifWritingAuxFile\immediate\write\auxfile{\noexpand\xdef\noexpand#1{#1}}\fi%
}
\def\appendix{\Appendix}
\def\Subappendix#1{\global\advance\subsecno by1\relax\medskip %
\leftline{\bf\romappno.\the\subsecno\ #1}%
\par\nobreak\smallskip\nobreak}
\def\tagsubappendix#1{\makeNormalizedRomappno%
\warnIfChanged#1{\normalizedromappno.\the\subsecno}%
\xdef#1{\normalizedromappno.\the\subsecno}%
\ifWritingAuxFile\immediate\write\auxfile{\noexpand\xdef\noexpand#1{#1}}\fi%
}

\def\eqn#1{\makeNormalizedRomappno%
\ifnum\secno>0%
  \warnIfChanged#1{\the\secno.\the\meqno}%
  \eqno(\the\secno.\the\meqno)\xdef#1{\the\secno.\the\meqno}%
     \global\advance\meqno by1
\else\ifnum\appno>0%
  \warnIfChanged#1{\normalizedromappno.\the\meqno}%
  \eqno({\rm\romappno}.\the\meqno)%
      \xdef#1{\normalizedromappno.\the\meqno}%
     \global\advance\meqno by1
\else%
  \warnIfChanged#1{\the\meqno}%
  \eqno(\the\meqno)\xdef#1{\the\meqno}%
     \global\advance\meqno by1
\fi\fi%
\eqlabeL#1%
\ifWritingAuxFile\immediate\write\auxfile{\noexpand\xdef\noexpand#1{#1}}\fi%
}
\def\defeqn#1{\makeNormalizedRomappno%
\ifnum\secno>0%
  \warnIfChanged#1{\the\secno.\the\meqno}%
  \xdef#1{\the\secno.\the\meqno}%
     \global\advance\meqno by1
\else\ifnum\appno>0%
  \warnIfChanged#1{\normalizedromappno.\the\meqno}%
  \xdef#1{\normalizedromappno.\the\meqno}%
     \global\advance\meqno by1
\else%
  \warnIfChanged#1{\the\meqno}%
  \xdef#1{\the\meqno}%
     \global\advance\meqno by1
\fi\fi%
\eqlabeL#1%
\ifWritingAuxFile\immediate\write\auxfile{\noexpand\xdef\noexpand#1{#1}}\fi%
}
\def\anoneqn{\makeNormalizedRomappno%
\ifnum\secno>0
  \eqno(\the\secno.\the\meqno)%
     \global\advance\meqno by1
\else\ifnum\appno>0
  \eqno({\rm\normalizedromappno}.\the\meqno)%
     \global\advance\meqno by1
\else
  \eqno(\the\meqno)%
     \global\advance\meqno by1
\fi\fi%
}
\def\mfig#1#2{\ifx#20
\else\global\advance\figno by1%
\relax#1\the\figno%
\warnIfChanged#2{\the\figno}%
\xdef#2{\the\figno}%
\reflabeL#2%
\ifWritingAuxFile\immediate\write\auxfile{\noexpand\xdef\noexpand#2{#2}}\fi\fi%
}

\def\fig#1{\mfig{fig.\ ~}#1}

\catcode`@=11 

\newif\ifFiguresInText\FiguresInTexttrue
\newif\if@FigureFileCreated
\newwrite\capfile
\newwrite\figfile

\newif\ifcaption
\captiontrue
\def\captionsize{\tenrm}
\def\PlaceTextFigure#1#2#3#4{%
\vskip 0.5truein%
#3\hfil\epsfbox{#4}\hfil\break%
\ifcaption\hfil\vbox{\captionsize Figure #1. #2}\hfil\fi%
\vskip10pt}
\def\PlaceEndFigure#1#2{%
\epsfxsize=\hsize\epsfbox{#2}\vfill\centerline{Figure #1.}\eject}

\def\LoadFigure#1#2#3#4{%
\ifundef#1{\phantom{\mfig{}#1}}\else
\fi%
\ifFiguresInText
\PlaceTextFigure{#1}{#2}{#3}{#4}%
\else
\if@FigureFileCreated\else%
\immediate\openout\capfile=\jobname.caps%
\immediate\openout\figfile=\jobname.figs%
@FigureFileCreatedtrue\fi%
\immediate\write\capfile{\noexpand\item{Figure \noexpand#1.\ }{#2}\vskip10pt}%
\immediate\write\figfile{\noexpand\PlaceEndFigure\noexpand#1{\noexpand#4}}%
\fi}

\def\listfigs{\ifFiguresInText\else%
\vfill\eject\immediate\closeout\capfile
\immediate\closeout\figfile%
\centerline{{\bf Figures}}\bigskip\frenchspacing%
\catcode`@=11 
\def\captionsize{\tenrm}
\input \jobname.caps\vfill\eject\nonfrenchspacing%
\catcode`\@=\active
\catcode`@=12  
\input\jobname.figs\fi}

\font\ninerm=cmr9
\font\eightrm=cmr8
\font\sixrm=cmr6

\def\loadtrueseventeenpoint{
 \font\seventeenrm=cmr10 at 17.28truept
 \font\seventeeni=cmmi10 at 17.28truept
 \font\seventeenbf=cmbx10 at 17.28truept
 \font\seventeenit=cmti10 at 17.28truept
 \font\seventeensl=cmsl10 at 17.28truept
 \font\seventeensy=cmsy10 at 17.28truept
}
\def\loadfourteenpoint{
\font\fourteenrm=cmr10 at 14.4pt
\font\fourteeni=cmmi10 at 14.4pt
\font\fourteenit=cmti10 at 14.4pt
\font\fourteensl=cmsl10 at 14.4pt
\font\fourteensy=cmsy10 at 14.4pt
\font\fourteenbf=cmbx10 at 14.4pt
}
\def\loadtruetwelvepoint{
\font\twelverm=cmr10 at 12truept
\font\twelvei=cmmi10 at 12truept
\font\twelveit=cmti10 at 12truept
\font\twelvesl=cmsl10 at 12truept
\font\twelvesy=cmsy10 at 12truept
\font\twelvebf=cmbx10 at 12truept
}
\def\loadtwelvepoint{
\font\twelverm=cmr10 at 12pt
\font\twelvei=cmmi10 at 12pt
\font\twelveit=cmti10 at 12pt
\font\twelvesl=cmsl10 at 12pt
\font\twelvesy=cmsy10 at 12pt
\font\twelvebf=cmbx10 at 12pt
\font\elevenrm=cmr10 at 11pt
}
\font\ninei=cmmi9
\font\eighti=cmmi8
\font\sixi=cmmi6
\skewchar\ninei='177 \skewchar\eighti='177 \skewchar\sixi='177

\font\ninesy=cmsy9
\font\eightsy=cmsy8
\font\sixsy=cmsy6
\skewchar\ninesy='60 \skewchar\eightsy='60 \skewchar\sixsy='60

\font\ninebf=cmbx9
\font\eightbf=cmbx8
\font\sixbf=cmbx6

\font\ninett=cmtt9
\font\eighttt=cmtt8

\hyphenchar\tentt=-1 
\hyphenchar\ninett=-1
\hyphenchar\eighttt=-1

\font\ninesl=cmsl9
\font\eightsl=cmsl8

\font\nineit=cmti9
\font\eightit=cmti8


\newskip\ttglue
\def\tenpoint{\def\rm{\fam0\tenrm}%
  \textfont0=\tenrm \scriptfont0=\sevenrm \scriptscriptfont0=\fiverm
  \textfont1=\teni \scriptfont1=\seveni \scriptscriptfont1=\fivei
  \textfont2=\tensy \scriptfont2=\sevensy \scriptscriptfont2=\fivesy
  \textfont3=\tenex \scriptfont3=\tenex \scriptscriptfont3=\tenex
  \def\it{\fam\itfam\tenit}\textfont\itfam=\tenit
  \def\sl{\fam\slfam\tensl}\textfont\slfam=\tensl
  \def\bf{\fam\bffam\tenbf}\textfont\bffam=\tenbf \scriptfont\bffam=\sevenbf
  \scriptscriptfont\bffam=\fivebf
  \normalbaselineskip=12pt
  \let\sc=\eightrm
  \let\big=\tenbig
  \setbox\strutbox=\hbox{\vrule height8.5pt depth3.5pt width\z@}%
  \normalbaselines\rm}

\def\twelvepoint{\def\rm{\fam0\twelverm}%
  \textfont0=\twelverm \scriptfont0=\ninerm \scriptscriptfont0=\sevenrm
  \textfont1=\twelvei \scriptfont1=\ninei \scriptscriptfont1=\seveni
  \textfont2=\twelvesy \scriptfont2=\ninesy \scriptscriptfont2=\sevensy
  \textfont3=\tenex \scriptfont3=\tenex \scriptscriptfont3=\tenex
  \def\it{\fam\itfam\twelveit}\textfont\itfam=\twelveit
  \def\sl{\fam\slfam\twelvesl}\textfont\slfam=\twelvesl
  \def\bf{\fam\bffam\twelvebf}\textfont\bffam=\twelvebf%
  \scriptfont\bffam=\ninebf
  \scriptscriptfont\bffam=\sevenbf
  \normalbaselineskip=12pt
  \let\sc=\eightrm
  \let\big=\tenbig
  \setbox\strutbox=\hbox{\vrule height8.5pt depth3.5pt width\z@}%
  \normalbaselines\rm}

\def\fourteenpoint{\def\rm{\fam0\fourteenrm}%
  \textfont0=\fourteenrm \scriptfont0=\tenrm \scriptscriptfont0=\sevenrm
  \textfont1=\fourteeni \scriptfont1=\teni \scriptscriptfont1=\seveni
  \textfont2=\fourteensy \scriptfont2=\tensy \scriptscriptfont2=\sevensy
  \textfont3=\tenex \scriptfont3=\tenex \scriptscriptfont3=\tenex
  \def\it{\fam\itfam\fourteenit}\textfont\itfam=\fourteenit
  \def\sl{\fam\slfam\fourteensl}\textfont\slfam=\fourteensl
  \def\bf{\fam\bffam\fourteenbf}\textfont\bffam=\fourteenbf%
  \scriptfont\bffam=\tenbf
  \scriptscriptfont\bffam=\sevenbf
  \normalbaselineskip=17pt
  \let\sc=\elevenrm
  \let\big=\tenbig
  \setbox\strutbox=\hbox{\vrule height8.5pt depth3.5pt width\z@}%
  \normalbaselines\rm}

\def\seventeenpoint{\def\rm{\fam0\seventeenrm}%
  \textfont0=\seventeenrm \scriptfont0=\fourteenrm \scriptscriptfont0=\tenrm
  \textfont1=\seventeeni \scriptfont1=\fourteeni \scriptscriptfont1=\teni
  \textfont2=\seventeensy \scriptfont2=\fourteensy \scriptscriptfont2=\tensy
  \textfont3=\tenex \scriptfont3=\tenex \scriptscriptfont3=\tenex
  \def\it{\fam\itfam\seventeenit}\textfont\itfam=\seventeenit
  \def\sl{\fam\slfam\seventeensl}\textfont\slfam=\seventeensl
  \def\bf{\fam\bffam\seventeenbf}\textfont\bffam=\seventeenbf%
  \scriptfont\bffam=\fourteenbf
  \scriptscriptfont\bffam=\twelvebf
  \normalbaselineskip=21pt
  \let\sc=\fourteenrm
  \let\big=\tenbig
  \setbox\strutbox=\hbox{\vrule height 12pt depth 6pt width\z@}%
  \normalbaselines\rm}

\def\ninepoint{\def\rm{\fam0\ninerm}%
  \textfont0=\ninerm \scriptfont0=\sixrm \scriptscriptfont0=\fiverm
  \textfont1=\ninei \scriptfont1=\sixi \scriptscriptfont1=\fivei
  \textfont2=\ninesy \scriptfont2=\sixsy \scriptscriptfont2=\fivesy
  \textfont3=\tenex \scriptfont3=\tenex \scriptscriptfont3=\tenex
  \def\it{\fam\itfam\nineit}\textfont\itfam=\nineit
  \def\sl{\fam\slfam\ninesl}\textfont\slfam=\ninesl
  \def\bf{\fam\bffam\ninebf}\textfont\bffam=\ninebf \scriptfont\bffam=\sixbf
  \scriptscriptfont\bffam=\fivebf
  \normalbaselineskip=11pt
  \let\sc=\sevenrm
  \let\big=\ninebig
  \setbox\strutbox=\hbox{\vrule height8pt depth3pt width\z@}%
  \normalbaselines\rm}

\def\eightpoint{\def\rm{\fam0\eightrm}%
  \textfont0=\eightrm \scriptfont0=\sixrm \scriptscriptfont0=\fiverm%
  \textfont1=\eighti \scriptfont1=\sixi \scriptscriptfont1=\fivei%
  \textfont2=\eightsy \scriptfont2=\sixsy \scriptscriptfont2=\fivesy%
  \textfont3=\tenex \scriptfont3=\tenex \scriptscriptfont3=\tenex%
  \def\it{\fam\itfam\eightit}\textfont\itfam=\eightit%
  \def\sl{\fam\slfam\eightsl}\textfont\slfam=\eightsl%
  \def\bf{\fam\bffam\eightbf}\textfont\bffam=\eightbf \scriptfont\bffam=\sixbf%
  \scriptscriptfont\bffam=\fivebf%
  \normalbaselineskip=9pt%
  \let\sc=\sixrm%
  \let\big=\eightbig%
  \setbox\strutbox=\hbox{\vrule height7pt depth2pt width\z@}%
  \normalbaselines\rm}

\def\tenbig#1{{\hbox{$\left#1\vbox to8.5pt{}\right.\n@space$}}}
\def\ninebig#1{{\hbox{$\textfont0=\tenrm\textfont2=\tensy
  \left#1\vbox to7.25pt{}\right.\n@space$}}}
\def\eightbig#1{{\hbox{$\textfont0=\ninerm\textfont2=\ninesy
  \left#1\vbox to6.5pt{}\right.\n@space$}}}

\def\footnote#1{\edef\@sf{\spacefactor\the\spacefactor}#1\@sf
      \insert\footins\bgroup\eightpoint
      \interlinepenalty100 \let\par=\endgraf
        \leftskip=\z@skip \rightskip=\z@skip
        \splittopskip=10pt plus 1pt minus 1pt \floatingpenalty=20000
        \smallskip\item{#1}\bgroup\strut\aftergroup\@foot\let\next}
\skip\footins=12pt plus 2pt minus 4pt 
\dimen\footins=30pc 

\newinsert\margin
\dimen\margin=\maxdimen

\loadtruetwelvepoint 
\loadtrueseventeenpoint

\def\eatOne#1{}
\def\ifundef#1{\expandafter\ifx%
\csname\expandafter\eatOne\string#1\endcsname\relax}
\def\notTrue{\iffalse}\def\isTrue{\iftrue}
\def\ifdef#1{{\ifundef#1%
\aftergroup\notTrue\else\aftergroup\isTrue\fi}}
\def\use#1{\ifundef#1\linemessage{Warning: \string#1 is undefined.}%
{\tt \string#1}\else#1\fi}



%
\catcode`"=11
\let\quote="
\catcode`"=12
\chardef\foo="22
\global\newcount\refno \global\refno=1
\newwrite\rfile
\newlinechar=`\^^J
\def\@ref#1#2{\the\refno\n@ref#1{#2}}
\def\h@ref#1#2#3{\href{#3}{\the\refno}\n@ref#1{#2}}
\def\n@ref#1#2{\xdef#1{\the\refno}%
\ifnum\refno=1\immediate\openout\rfile=\jobname.refs\fi%
\immediate\write\rfile{\noexpand\item{[\noexpand#1]\ }#2.}%
\global\advance\refno by1}
\def\nref{\n@ref} 
\def\ref{\@ref}   
\def\hrref{\h@ref}
\def\lref#1#2{\the\refno\xdef#1{\the\refno}%
\ifnum\refno=1\immediate\openout\rfile=\jobname.refs\fi%
\immediate\write\rfile{\noexpand\item{[\noexpand#1]\ }#2\semi}%
\global\advance\refno by1}
\def\cref#1{\immediate\write\rfile{#1\semi}}

\def\preref#1#2{\gdef#1{\@ref#1{#2}}}

\def\semi{;\hfil\noexpand\break}

\def\listrefs{\vfill\eject\immediate\closeout\rfile
\centerline{{\bf References}}\bigskip\frenchspacing%
\input \jobname.refs\vfill\eject\nonfrenchspacing}

\def\inputAuxIfPresent#1{\immediate\openin1=#1
\ifeof1\message{No file \auxfileName; I'll create one.
}\else\closein1\relax\input\auxfileName\fi%
}




\newif\ifWritingAuxFile
\newwrite\auxfile
\def\SetUpAuxFile{%
\xdef\auxfileName{\jobname.aux}%
\inputAuxIfPresent{\auxfileName}%
\WritingAuxFiletrue%
\immediate\openout\auxfile=\auxfileName}



\catcode`\@=\active
\catcode`@=12  
\catcode`\"=\active


\def\deltatheory{(\Phi^2)^{2-\delta}}
\def\loopint{\int {d^{d}l \over (2\pi)^d} } \def\half{{1\over 2}} \def\c{\cdot}
\def\del{\partial}

\def\zofj{ {Z\left[ J \right] = \int \left[ d\Phi \right] {\rm exp}
\Bigl( { i \langle\half\partial_{\alpha} \Phi \partial^{\alpha} \Phi -
\half m^2
\Phi^2 - {\lambda \over 4!} \deltatheory + J \Phi \rangle } }
\Bigr) }

\newread\epsffilein    
\newif\ifepsffileok    
\newif\ifepsfbbfound   
\newif\ifepsfverbose   
\newdimen\epsfxsize    
\newdimen\epsfysize    
\newdimen\epsftsize    
\newdimen\epsfrsize    
\newdimen\epsftmp      
\newdimen\pspoints     
\pspoints=1bp          
\epsfxsize=0pt         
\epsfysize=0pt         
\def\epsfbox#1{\global\def\epsfllx{72}\global\def\epsflly{72}%
   \global\def\epsfurx{540}\global\def\epsfury{720}%
   \def\lbracket{[}\def\testit{#1}\ifx\testit\lbracket
   \let\next=\epsfgetlitbb\else\let\next=\epsfnormal\fi\next{#1}}%
\def\epsfgetlitbb#1#2 #3 #4 #5]#6{\epsfgrab #2 #3 #4 #5 .\\%
   \epsfsetgraph{#6}}%
\def\epsfnormal#1{\epsfgetbb{#1}\epsfsetgraph{#1}}%
\def\epsfgetbb#1{%
%
%
\openin\epsffilein=#1
\ifeof\epsffilein\errmessage{I couldn't open #1, will ignore it}\else
%
%
   {\epsffileoktrue \chardef\other=12
    \def\do##1{\catcode`##1=\other}\dospecials \catcode`\ =10
    \loop
       \read\epsffilein to \epsffileline
       \ifeof\epsffilein\epsffileokfalse\else
%
%
          \expandafter\epsfaux\epsffileline:. \\%
       \fi
   \ifepsffileok\repeat
   \ifepsfbbfound\else
    \ifepsfverbose\message{No bounding box comment in #1; using defaults}\fi\fi
   }\closein\epsffilein\fi}%
%
%
\def\epsfclipstring{}
\def\epsfsetgraph#1{%
   \epsfrsize=\epsfury\pspoints
   \advance\epsfrsize by-\epsflly\pspoints
   \epsftsize=\epsfurx\pspoints
   \advance\epsftsize by-\epsfllx\pspoints
%
%
   \epsfxsize\epsfsize\epsftsize\epsfrsize
   \ifnum\epsfxsize=0 \ifnum\epsfysize=0
      \epsfxsize=\epsftsize \epsfysize=\epsfrsize
      \epsfrsize=0pt
%
%
     \else\epsftmp=\epsftsize \divide\epsftmp\epsfrsize
       \epsfxsize=\epsfysize \multiply\epsfxsize\epsftmp
       \multiply\epsftmp\epsfrsize \advance\epsftsize-\epsftmp
       \epsftmp=\epsfysize
       \loop \advance\epsftsize\epsftsize \divide\epsftmp 2
       \ifnum\epsftmp>0
          \ifnum\epsftsize<\epsfrsize\else
             \advance\epsftsize-\epsfrsize \advance\epsfxsize\epsftmp \fi
       \repeat
       \epsfrsize=0pt
     \fi
   \else \ifnum\epsfysize=0
     \epsftmp=\epsfrsize \divide\epsftmp\epsftsize
     \epsfysize=\epsfxsize \multiply\epsfysize\epsftmp
     \multiply\epsftmp\epsftsize \advance\epsfrsize-\epsftmp
     \epsftmp=\epsfxsize
     \loop \advance\epsfrsize\epsfrsize \divide\epsftmp 2
     \ifnum\epsftmp>0
        \ifnum\epsfrsize<\epsftsize\else
           \advance\epsfrsize-\epsftsize \advance\epsfysize\epsftmp \fi
     \repeat
     \epsfrsize=0pt
    \else
     \epsfrsize=\epsfysize
    \fi
   \fi
%
%
   \ifepsfverbose\message{#1: width=\the\epsfxsize, height=\the\epsfysize}\fi
   \epsftmp=10\epsfxsize \divide\epsftmp\pspoints
   \vbox to\epsfysize{\vfil\hbox to\epsfxsize{%
      \ifnum\epsfrsize=0\relax
        \includegraphics{#1}%
      \else
        \epsfrsize=10\epsfysize \divide\epsfrsize\pspoints
        \includegraphics{#1}%
      \fi
      \hfil}}%
\global\epsfxsize=0pt\global\epsfysize=0pt}%
%
%
{\catcode`\%=12 \global\let\epsfpercent=
%
%
\long\def\epsfaux#1#2:#3\\{\ifx#1\epsfpercent
   \def\testit{#2}\ifx\testit\epsfbblit
      \epsfgrab #3 . . . \\%
      \epsffileokfalse
      \global\epsfbbfoundtrue
   \fi\else\ifx#1\par\else\epsffileokfalse\fi\fi}%
%
%
\def\epsfempty{}%
\def\epsfgrab #1 #2 #3 #4 #5\\{%
\global\def\epsfllx{#1}\ifx\epsfllx\epsfempty
      \epsfgrab #2 #3 #4 #5 .\\\else
   \global\def\epsflly{#2}%
   \global\def\epsfurx{#3}\global\def\epsfury{#4}\fi}%
%
%
\def\epsfsize#1#2{\epsfxsize}
%
%


\preref\BenderOne{ Carl M.\ Bender, Kimball A.\ Milton, Moshe Moshe, Stephen,
L.M.\ Simmons, Jr., Phys.\ Rev.\ D37:1472 (1988)}

\preref\BenderTwo{ Carl M.\ Bender, Kimball A.\ Milton, Stephen S.\ Pinsky,
L.M.\
Simmons, Jr., J.\ Math.\ Phys.\ 31:2722-2725 (1990)}

\preref\BenderThree{ Stephen S.\ Pinsky, L.M.\ Simmons, Jr.,
Phys.Rev.D38:2518, (1988)}

\preref\triviality{ R.\ Fernandez, J.\ Frohlich, A.D.\ Sokal, Springer (1992)
444 pp}

\preref\Bateman{ Bateman Manuscript Project, Volume III, Mc-Graw Hill Book
Co., (1955)}

\preref\irr{ Quantum Field Theory and Critical Phenomena , J.\ Zinn-Justin,
Oxford, UK: Clarendon Press (1989)}

\preref\polch{ Joseph Polchinski, Nucl.\ Phys.\ B231:269-295 (1984)}

\preref\ch{ Gordon Chalmers, Phys.\ Rev.\ D53:7143-7156 (1996)}

\preref\soon{ Gordon Chalmers, in preparation}


\loadtwelvepoint
\baselineskip=24pt

\def\eps{\epsilon}
\def\c{\cdot}
\def\half{{1\over 2}}


\rightline{ITP-SB-73}
\rightline{hep-th yymmdd}

\vglue 1.0 cm
\centerline{\bf Quantization of Non-Polynomial Field Theories}

\vglue 1cm
\centerline{\bf Gordon Chalmers} \vskip .2 cm
\baselineskip=13pt
\centerline{\it Institute for Theoretical Physics}
\centerline{\it State  University of New York}
\centerline{\it Stony Brook, NY 11794-3840}
\centerline{\it chalmers@insti.physics.sunysb.edu}
\vglue 1cm

\centerline{\bf ABSTRACT}
\vglue 0.3cm

We re-examine the quantization of a class of non-polynomial scalar  field
theories which
interpolates continuously from a free one to $\phi^4$ theory. The
quantization of such
theories is problematic because the Feynman  rules may not be directly
obtained. We give a
means for calculating the correlation functions in this theory. The Feynman
rules
developed here shall enable further progress in the understanding of the
triviality of
$\phi^4$ theory in four dimensions.

\vfill
\eject


\baselineskip=15pt
\loadfourteenpoint

\vskip .3in
\section{Introduction}
\vskip .3in

In previous years, a novel perturbative scheme was found enabling  one to
calculate
correlation functions in certain non-polynomial scalar  field theories. The
interaction is
taken to be $\deltatheory$  [\use\BenderOne, \use\BenderTwo,
\use\BenderThree], and gives a non-polynomial theory containing logarithms,
$$
\eqalign{
\deltatheory & = \phi^4 e^{-\delta \ln \phi^2} \cr & =\phi^4 -
\delta \phi^4 \ln \phi^2 + \ldots } \ . \eqn\dtheory
$$ Such a theory continuously interpolates between a free theory  and the
standard
$\phi^4$ scalar model.

There are a few interesting aspects of a theory with this potential.
Perhaps  the most
intriguing one pertains to the fact that $\phi^4$ theory in four dimensions
may actually
be non-interacting, although no analytical  evidence for this has been
found. This is the
so-called triviality  problem in four dimensions which to this day still
remains
unsettled,  despite strong numerical evidence. To ask whether a
$\phi^4$ theory remains  trivially interacting when coupled to other
fields, such as a
non-abelian  gauge theory, is a different question. However, in light of
the fact that
the Higgs mechanism is the foundation of the electroweak sector it is an
important
question to address.

Varying the parameter $\delta$ in the theory with the
interaction above in eq.~(\use\dtheory) allows one to interpolate from
$\delta=1$ along a path $\phi^4 \vert \phi\vert^{-2\delta}$ and  explore the
continuous relationship between the dimension of space-time
$d$ and the interaction parameter $\delta$. The scalar field theory with a
potential
$\phi^4$ is expected to be interacting in $d<4$, where the coupling obtains a
positive mass scale. Intuitively, we expect to find non-trivial
interactions when the
coupling constant has a positive mass scale, so that the behavior of the
full field
theory should be tied strongly to both $d$ and $\delta$. Indeed, the triviality
question can be phrased in the context of whether or not the $\phi^4$
operator in the
full quantum field theory may acquire scaling dimensions away from four, which
effectively generates a scale for the coupling. The analysis of Greens
functions in
the space of values $(d,\delta)$ sheds further insight into the triviality
problem.

Further issues addressed in the study of this theory concerns the behavior
of field
theories containing an infinite number of monomial interactions. Within the
analysis
of the renormalization group we know that to any finite order in
perturbation theory,
only a few operators, namely the relevant ones, dominate the behavior of
the Greens
functions at low energy. Corrections arising from higher-dimensional
interactions,
and which are suppressed by the cutoff used to regularize the theory, are
suppressed
by powers of the renormalization scale. In the case of scalar theories in four
dimensions the mass operator $\phi^2$ is the only relevant one, while the
interaction
$\phi^4$ with classically zero scaling dimensions may or may not be: it is
a marginal
operator which perturbatively acquires non-vanishing scaling dimensions via
quantum
effects. As we will see, the theory with an interaction above may be
treated as a
theory with an infinite number of interactions simulating the non-polynomial
potential. Taking into account the effects of the infinite tower of couplings
remarkably changes the character of the Greens functions. The issues of small
perturbations of the theory, and the combined effect of a tower of irrelevant
interactions, is addressed in the context of the non-polynomial theory.

The first difficulty in working with such a theory is how to formulate a
perturbative
expansion of the Greens functions; the potential is not differentiable at
$\phi = 0$
and hence does not admit a Taylor expansion. There are no textbook Feynman
rules
present. One may resort to letting the mass parameter $m^2$ become
negative, followed
by an expansion about the broken phase from which the scalar field acquires a
non-vanishing vacuum value. Another option, which we develop in these
notes, is to
find an analytic continuation in $\delta$ in which the Greens functions may be
defined. We will see that there are several different means of defining
this theory,
all of which lead to the same perturbative results.

This work is broken into several sections. In section 2, we develop the
perturbative
means for calculating the Greens functions. In Section 3 we perform the
next-to-lowest order (in $\lambda$) calculation of the four-point function and
investigate the behavior as a function of the dimension and interaction
parameter.

\vskip .3in
\section{Perturbative Expansion}
\vskip .2in

We take our theory to be a free massive scalar theory coupled to an
interaction of
the form

$$ {\cal L}_{\rm int} = {\lambda\over 4} \deltatheory \ ,
\anoneqn
$$ and are interested both in exploring the behavior of the theory as a
function of
the dimension $d$, as well as in the possible momentum cutoff $\Lambda$. In $d$
dimensions, the mass scales of the fields and coupling constants are
$$ [\phi] = {d\over 2} - 1
$$
$$ [\lambda ] = 2\delta +(d-4) (1-\delta ) \ ,
\anoneqn
$$ and the field theory is naively power counting renormalizable for $$ d < 4 +
{2\delta \over 1-\delta} \ .
\anoneqn
$$
\vskip .2in

\noindent Note that even in four dimensions the coupling has a mass scale when
$\delta$ is not zero. The space of naive power-counting renormalizable
theories, in
which
$[\lambda ] \geq 0$, is illustrated in \fig\ShadedRegionFigure\ as a
function of the
dimension and the interaction parameter $\delta$.

\vskip -.6 cm
\LoadFigure\ShadedRegionFigure{\baselineskip 13 pt
\noindent\narrower\ninerm  The
shaded region denotes the parameter space of $(d,\delta)$ leading to  naive
power-counting renormalizability.} {\epsfysize 3truein }{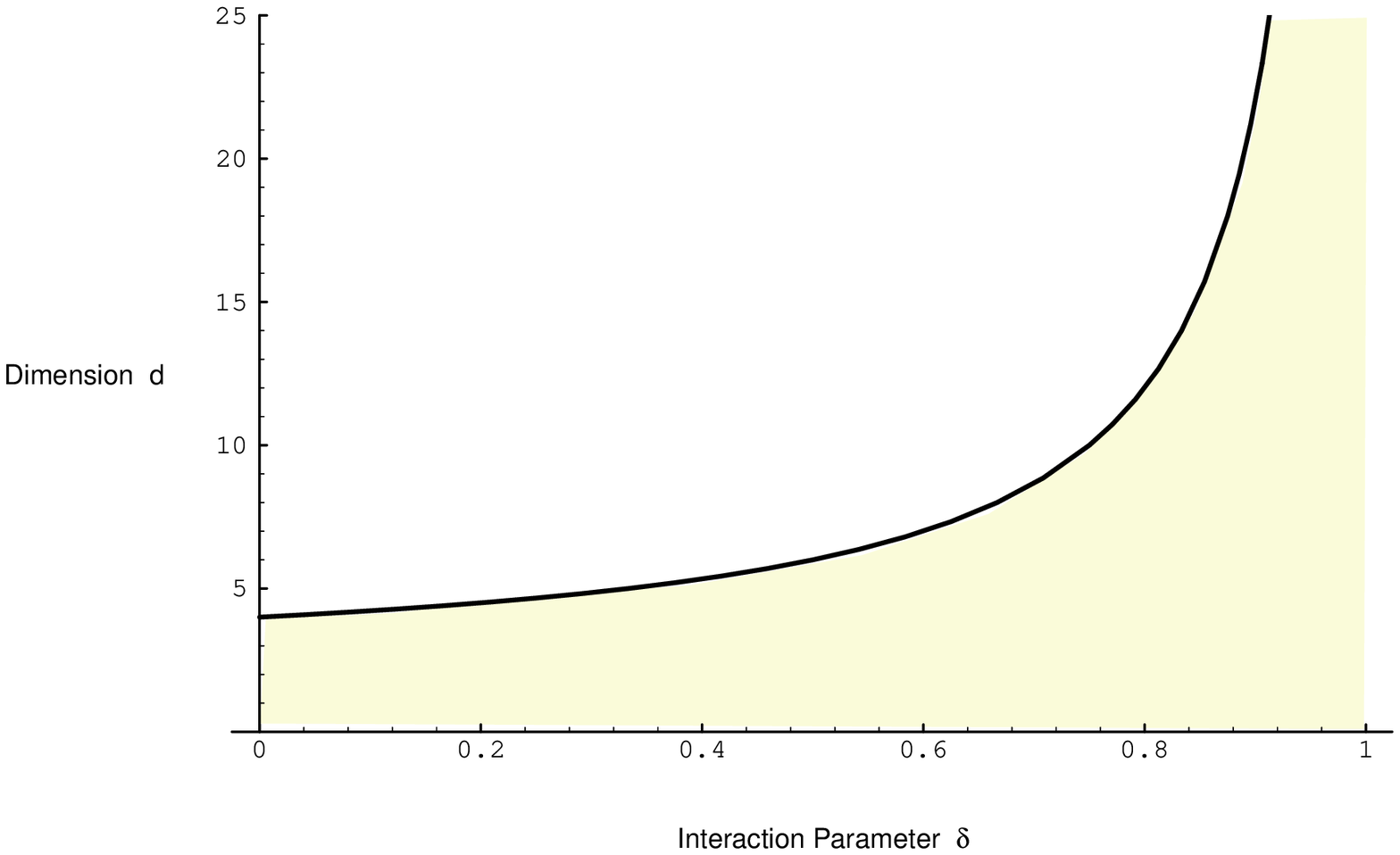}

\vskip .2in

For general values of $d$ and $\delta$ the coupling constant $\lambda$
becomes a
classical mass scale. We extract the arbitrary scale $\mu$ to keep the coupling
constant $\lambda$ dimensionless; the full Lagrangian in $d=4-\eps$
dimensions is then

$$ {\cal L}_{d} = \half\del\phi\del\phi - \half m^2 \phi^2 -
\mu^{2\delta+\eps (1-\delta)} {\lambda\over 4!} \deltatheory
\eqn\dimreglagrangian
$$ We will drop for now the $\mu$ scale and re-insert it later.

In the second formulation we define the theory in Euclidean space with a
momentum
cutoff
$\Lambda$, thus explicitly breaking Poincare invariance. With this
regularization the
dimension of the coupling constant by the criteria of naturalness [\use\polch ]
should be replaced by the appropriate power of the cutoff $\Lambda$. The
Lagrangian
becomes

$$ {\cal L}_\Lambda = \half\del\phi\del\phi + \half m^2 \phi^2 +
\Lambda^{2\delta}
{\lambda\over 4!} \deltatheory
\eqn\cutofflagrangian
$$ Again we drop the $\Lambda^{2\delta}$ for now and insert it back later.
In the more
general case we may give the dimensions $(\Lambda^{2- a} m^a)^\delta$ to
the coupling
constant (where
$2\geq a \geq 0$). The details of deriving the effective Feynman rules do
not depend
on the two choices of the regularization schemes.

\vskip .2in
\subsection{Perturbative Rules}

In this section we develop the formalism for doing perturbative
calculations despite
the non-Taylor expandable interaction in (\use\dtheory). In the presence of
a source
term ${\cal L}=\int J\phi$, the connected Greens functions are found by the
appropriate functional derivatives of $Z \left[ J \right]$ with respect to
the source
$J(x_i)$. The path integral is defined as

$$
\zofj \ ,
\anoneqn
$$
>from which we derive the Greens functions

$$
\eqalign{ G^{(n)} (x_1,x_2,\ldots,x_n) = & (-i)^n {\delta \over \delta J(x_1)}
{\delta \over
\delta J(x_2)} \cdots {\delta
\over \delta J(x_n)} \ln Z \left[ J \right] \vert_{J=0} \cr & = \langle
\phi (x_1) \cdots
\phi (x_n) \rangle \ . } \anoneqn
$$ For polynomial field theories the standard means of generating the
connected Greens
functions is derived by explicitly taking the functional derivatives of
$\ln Z \left[ J
\right]$ and applying Wicks theorem in the expansion of the functional
integral.
However, the potential in question is not polynomial and there is no analog
of Wick's
theorem that disentangles a non-integer number of fields. Clearly then
there are no
Feynman rules in the standard sense for this theory.

The first course of action we take is the following:  We perform an
expansion of the
potential in terms of Laguerre polynomials (given by a sum of monomials)
followed by
a resummation of all of the self-energy corrections. In effect  this
converts the original
non-polynomial potential into an infinite sum of interactions, all of which are
normal-ordered by construction.

The generalized Laguerre polynomials [\use\Bateman] are

$$ L_n^{\alpha}(x) = \sum_{k=0}^n
{\Gamma(n+\alpha+1)\over\Gamma(k+\alpha+1)} {(-x)^k
\over k! (n-k)! } \ .
\anoneqn
$$

\noindent Note that we can replace the upper limit with $\infty$ since the
argument
of the sum vanishes for $k>n$. The above polynomials satisfy the
generalized Laguerre
differential equation. Next we use an expansion which is uniformly
convergent for all
$\phi^2 > 0$ and $-\delta > -{1\over 2}(\alpha +1)$:

$$ (\phi^2)^{-\delta} = \Gamma(1+\alpha-\delta) \Gamma(1-\delta)
\sum_{n=0}^{\infty}
{(-)^n L_n^{\alpha} (\phi^2) \over
\Gamma(1+\alpha+n) \Gamma(1-\delta-n) } \ . \eqn\lexpansion
$$

\noindent After inserting the Laguerre expansion into (\use\lexpansion) and
using the
Gamma function reflection identity we obtain the double infinite sum
representing the
interaction

$$ (\phi^2)^{-\delta} = - \Gamma(1+\alpha-\delta) \Gamma(1-\delta)
({\sin\pi\delta
\over \pi})
\sum_{n=0}^{\infty}
\sum_{k=0}^{\infty} {\Gamma(n+\delta) \over \Gamma(1+\alpha+k)} {
(-\phi^2)^{k} \over
k! (n-k)! }  \ .
\eqn\drexpansion
$$

\noindent At this point we have more or less formally manipulated the
non-polynomial
interaction and rewritten it as an infinite sum of polynomials. It is
important to
note that this (non-Taylor) expansion is uniformly convergent in the range
$\phi^2
>0$.

Using the expansion (\use\drexpansion) we can read off the Feynman rules.
Denote  by
$\lambda_{2k+4}$ the value of the $2k$-point tree vertex. Then the
potential is

$$ V(\phi) = \sum_{k=0}^{\infty} \lambda_{2k+4} {\phi^{2k+4} \over (2k+4)!} \ ,
\anoneqn
$$ where

$$
\lambda_{2k+4} = \lambda_{\alpha} (-)^{k+1} {\Gamma(2k+5) \over
\Gamma(\alpha +k+1)
\Gamma(k+1) }
\sum_{n=0}^{\infty} {\Gamma(n+\delta) \over \Gamma (1+n-k) } \ ,
\eqn\naivevertex
$$ and
$$
\lambda_{\alpha} = \lambda ~\Gamma(1+\alpha-\delta) \Gamma(1-\delta)
({\sin\pi\delta
\over \pi}) \ .
\eqn\vertices
$$ The extra factor of $\Gamma(2k+5)$ has been divided out of the
definition of the
coupling
$\lambda_{2k+4}$ to make up for the combinatorical factor associated with
Wick's
theorem. However, the couplings $\lambda_{2k+4}$ do not make sense because
the sums
in (\use\vertices) do not converge; upon summing over all self-interactions,
illustrated in \fig\DaisyFigure\ , we shall
derive the full vertices. The resummation effectively normal-orders all of the
polynomial interactions in (\use\naivevertex).

\vskip -.4 cm
\LoadFigure\DaisyFigure{\baselineskip 13 pt \noindent\narrower\ninerm An
effective
$2p+4$-point vertex is found by self-contracting $2m$ lines into tadpoles
>from the
$2m+2p+4$-point vertex
$\lambda_{2p+2m+4}$.} {\epsfysize 2.1truein}{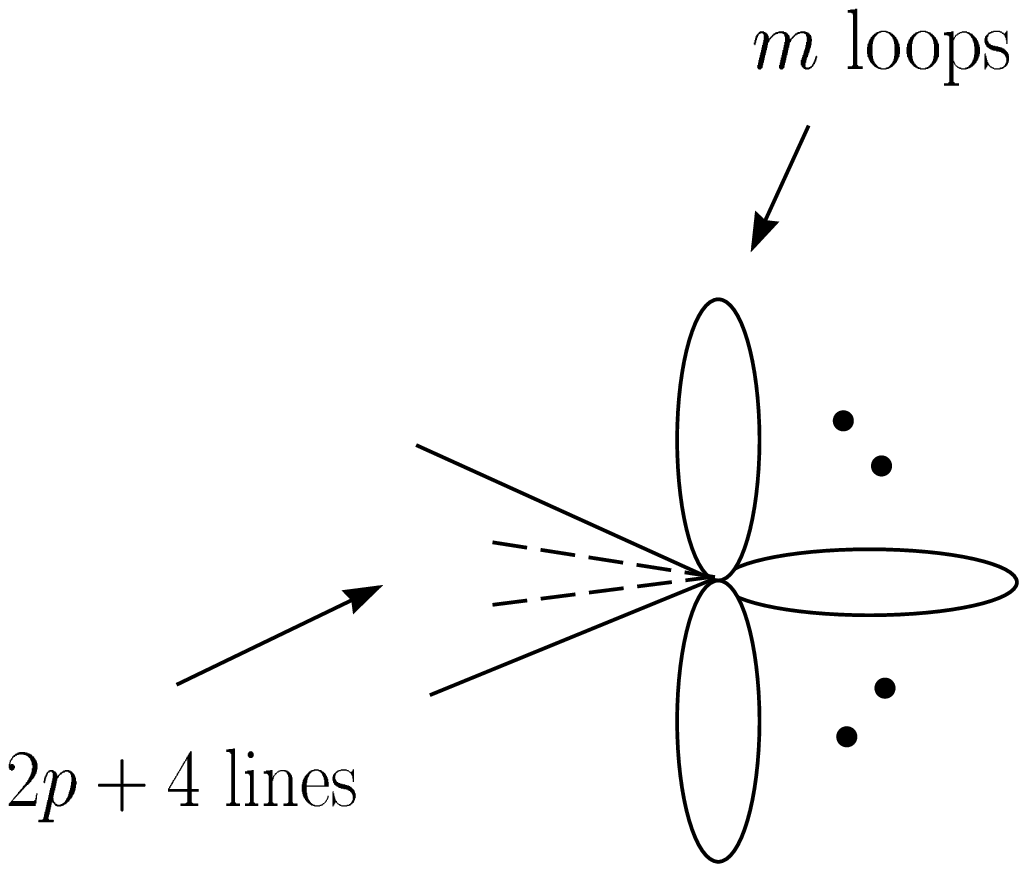}{}

\vskip .2in
If we sum over all of the
contributions we arrive at new zeroeth-order couplings $g_{2p}$ which by
definition
correspond to a normal-ordered (::) potential

$$
V(\phi) = \sum_{p=0}^{\infty} g_{2p} {:\phi^{2p}: \over (2p)!} \ .
\anoneqn
$$
We compute the sum over all daisy diagrams by taking the vertices in
eq.~(\use\naivevertex) and summing over all graphs formed by connecting
lines into
tad-pole configurations. In other words, a naive vertex with $2m+2p$ lines
can have
$2m$ of its lines contracted, thus forming $m$ tadpoles, leading to an
effective
vertex with $2p$ external lines. The graphs have a symmetry factor $1 \over
2^m m!$.
Summing over all the tadpole contractions of $2m$ lines from the vertices with
$2m+2p$ lines gives an infinite sum expression for the effective vertex
(where here
$p\geq -2$)\ ,

$$ g_{2p+4} = \sum_{m=0}^{\infty} \lambda_{2p+2m+4} ({I \over 2})^m {1\over
m!} \ .
\anoneqn
$$ The integral $I$ is defined to be the tadpole

$$ I \equiv \loopint {i \over l^2 - m^2} = \Gamma(1-{d\over 2})
(m^2)^{{d\over 2}-1}
(4\pi)^{-{d\over 2}} \ .
\eqn\tadpole
$$
\vskip .15in Adding together all of the contributions illustrated in
fig.~(\use\DaisyFigure ) then gives the infinite sum expression which
describes the
new coupling $g_{2p+4}$

$$ g_{2p+4} = \lambda_{\alpha} \sum_{n=0}^{\infty} \sum_{k=0}^{\infty}
{\Gamma(2p+2k+5)
\over
\Gamma(p+k+\alpha+1)
\Gamma(p+k+1) } {\Gamma(n+\delta) \over \Gamma(k+1) \Gamma(n-p-k+1)}
({I\over 2})^k
(-)^{p+k} \ .
\eqn\crap
$$
\vskip .2in
\noindent The evaluation of the sums in (\use\crap) may be performed as
follows:
first we shift $k$ by $k\rightarrow k-p$ since the inverse gamma function
contributes
zero for $k<p$. Using the identity
\vskip .2in
$$ {\Gamma(2k+5) \over \Gamma(k-p+3)} = 2^{2k+4} \pi^{-\half} ({\partial \over
\partial
\beta})^{p} \beta^{k+2}
\int^{\infty}_{-\infty} dt e^{-t^2} (t^2)^{k+2} \vert_{\beta=1} \ ,
\eqn\nicegamma
$$
\vskip .2in
\noindent we may rewrite the effective coupling in eq.~(\use\nicegamma) as

$$
\eqalign{ g_{2p} = & 16 \lambda_{\alpha}~ \pi^{-\half} ({I\over2})^{2-p}
({\partial
\over \partial
\beta})^p \beta^2 \int^{\infty}_{-\infty} dt ~ t^4 e^{-t^2} \cr & \c
\sum_{n=0}^{\infty} \sum_{k=0}^{\infty} {\Gamma(n+\delta) \over
\Gamma(k+1+\alpha)}
{1\over
\Gamma(k+1) \Gamma(n-k+1)} (-2 I \beta t^2)^k \vert_{\beta=1} \ . } \anoneqn
$$
\vskip .2in

\noindent We have interchanged the order of the sum and integration, which
is similar
to performing an analytic continuation in the expression in the parameter
$\delta$.
Performing the summation gives the compact result,

$$ g_{2p} = 16 \lambda_\alpha ~\pi^{-\half} ({I\over2})^{2-p} ({\partial \over
\partial \beta})^p
\beta^2
\int^{\infty}_{-\infty} dt ~t^4 e^{-t^2} \lambda (2 \beta t^2 I)^{-\delta} \ .
\anoneqn
$$
\vskip .2in

\noindent As a final step, the integral over $t$ and the derivatives with
respect to
$\beta$ are completely split and may be trivially evaluated: we arrive at
the result

$$ g_{2p} = \lambda_\alpha {\Gamma(5-2\delta) \over \Gamma(3-\delta-p)}
({I\over2})^{2-p-\delta}
\qquad p\geq 0~ \ ,
\eqn\vertices
$$
\vskip .2in
\noindent whereby in the derivation of the couplings $g_{2p}$ in
(\use\vertices) we
have effectively normal ordered the operators $\phi^{2p}$. The initial
Lagrangian
with the interaction $\deltatheory$ has been effectively redefined as an
infinite sum
of polynomial interactions

$$ L = \half \partial \phi \partial \phi - \half m^2 \phi^2 -
\sum_{p=0}^{\infty} {
g_{2p} \over (2p)! } : \phi^{2p} :  \ .
\eqn\normordlag
$$
Lastly, the effective theory above is constructed to {\it all orders} in
the free
parameter $\delta$. One can check that the usual free theory and $\phi^4$
theory are
obtained in the limits $\delta =1$ and $0$, respectively. Furthermore, at
tree-level
all of the interactions in eq.(\use\normordlag) at $2n\geq 4$-point vanish
due to the
suppression factor $I^{2-p-\delta}$ within the couplings $g_{2p}$.

One of the main motivations for the study of this theory, besides being a
means for dealing with scalar effective actions containing logarithms, was
to study
the triviality problem in four dimensions.  In future work we shall present
various loop calculations and analytical evidence [\use\soon ].

\vskip .3in
\noindent{\bf Conclusions}
\vskip .2in

In this work we have re-examined the quantization of a class of field
theories with
an interaction of the form $(\phi^2)^{2-2/\delta}$. Although the theory
does not
admit the usual perturbative definition of the Greens functions, we have
mapped the
interaction to a scalar theory containing an infinite number of polynomial
terms. The
calculation of Greens functions then follow as usual from the perturbative
rules.
In subsequent work we shall use our prescription for defining the
non-polynomial
theory to give some of the first analytical evidence supporting the triviality
of $\phi^4$ theory in four dimensions.

\listrefs

\vfill\break

\end